\documentclass[twocolumn,showpacs,aps,prb,showkeys,groupedaddress]{revtex4}
\usepackage{graphicx}
\usepackage{dcolumn}
\usepackage{bm}
\usepackage{amsmath}
\begin{document}

\title{Substrate-induced antiferromagnetism of an Fe monolayer on the Ir(001) surface}

\author{Josef Kudrnovsk\'y$^{1,2}$}\email{kudrnov@fzu.cz}
\author{Franti\v{s}ek M\'aca$^{1,2}$}
\author{Ilja Turek$^{3}$}
\author{Josef Redinger$^{4}$}
\affiliation{$^{1}$Institute of Physics ASCR, Na Slovance 2, CZ-182 21 Praha 8, Czech Republic\\
$^{2}$Max-Planck Institut f\"ur Mikrostrukturphysik,
Weinberg 2, D-06120 Halle, Germany\\
$^{3}$Institute of Physics of Materials ASCR, \v{Z}i\v{z}kova 22, CZ-616 62 Brno, Czech Republic\\
$^{4}$Department of General Physics, Vienna University of Technology,
Getreidemarkt 9/134, 1060 Vienna, Austria}


\newpage

\begin{abstract}

We present detailed ab initio study of structural and magnetic stability of a Fe-monolayer on the fcc(001) surface of
iridium. The Fe-monolayer has a strong tendency to order antiferromagnetically
for the true relaxed geometry.
On the contrary an unrelaxed Fe/Ir(001) sample has a
ferromagnetic ground state.
The antiferromagnetism is thus stabilized by
the decreased Fe-Ir layer spacing in striking contrast to the recently
experimentally observed antiferromagnetism of the
Fe/W(001) system which
exists also for an ideal bulk-truncated, unrelaxed geometry.
The calculated layer relaxations for Fe/Ir(001) agree reasonably well with
recent experimental LEED data.
The present study centers around the evaluation of
pair exchange interactions between Fe-atoms in the Fe-overlayer
as a function of the Fe/Ir  interlayer distance which
allows for a detailed understanding of the antiferromagnetism
of a Fe/Ir(001) overlayer.
Furthermore, our calculations indicate that the nature of the true
ground state could be more complex and display a
spin spiral-like rather than a c(2$\times $2)-antiferromagnetic order.
Finally, the magnetic stability of the Fe monolayer on the Ir(001)
surface is compared to the closely related Fe/Rh(001) system.

\end{abstract}

\pacs{75.30.-m,81.20.-n}
\keywords{surface magnetism, iron, iridium, density functional calculation}

\maketitle

\section{Introduction}

Magnetic overlayers, i.e., the thin films of magnetic materials
on the nonmagnetic substrate are systems with great technological
potential.
Magnetic overlayers are also a convenient system for a deeper
understanding of the origin of magnetism in the solid state.
The prototypical system is a magnetic monolayer (ML) on a non-magnetic
substrate which was the subject of many theoretical and experimental
studies in the past
\cite{chiral, martin, pajda, stepanyuk, hylgaard, freeman, Hafner, SB1, halle2, halle3, Ferriani, SB3, Hwang}.
Yet, a technological preparation, experimental study and detailed
theoretical understanding of the overlayer magnetism is still a
challenge to solid state physics.

There are a few important features which distinguish magnetic
overlayers from conventional magnetic materials.
This is, first of all, the presence of an external agent,
namely the presence of a non-magnetic substrate, which
can strongly modify the magnetic state and properties of
the overlayer system:
(i) The reduced coordination on the surface induces geometry as well as
 chemical binding changes.
(ii) Magnetic atoms adapt to the underlying lateral substrate lattice
spacings, particularly  true for a monolayer case.
(iii) The substrate electronic structure, in particular the position
of the substrate Fermi level is relevant for overlayer
magnetism.
(iv) The position of Fermi level can be tuned by the substrate electron
concentration, e.g. by growing the overlayer on an alloy
substrate with varying concentration of the constituents.
Alternatively, one could consider a substrate alloy which contains
magnetic and non-magnetic species or only  magnetic species, and/or
a partial coverage of the substrate
to vary the properties of the magnetic overlayer.
Theoretical and experimental studies on the above trends
help to establish a deeper understanding of the origin of surface magnetism.

First-principles calculations represent a powerful tool for
such studies, as they allow to determine reliably the
underlying lattice structure (possible layer relaxations,
surface reconstructions, etc.) which can be directly
compared with experiment (LEED).
They also allow to find out the underlying
magnetic structure although in this respect the situation is much
more complex. In addition to collinear magnetic
configurations (ferromagnetic (FM) and antiferromagnetic
(AFM) ones)  more complex configurations may exist (e.g.
recently observed chiral structures in bcc-Fe/W(001)
\cite{chiral}, magnetism of random overlayers \cite{martin},
etc.).
Consequently, an estimation and  discussion of  exchange interactions
 is very useful to gain a deeper understanding
of properties of both overlayer and bulk magnets and magnetic alloys, including
the diluted magnetic semiconductors \cite{ourdms,eirev}.
Surprisingly and in contrast to bulk systems, studies of exchange interactions
for magnetic overlayers are still very rare \cite{pajda,stepanyuk}
despite of their obvious importance.
In particular, the distance dependence of exchange integrals
can be very different from that in the bulk if, for example,
adatoms can interact via the host surface state as e.g. Co-adatoms
on the fcc(111) faces on noble metals \cite{hylgaard,stepanyuk}.

Great emphasis, both theoretical and experimental, has
been put recently on the study of the magnetic properties of
Fe-overlayers on the  (001) and (110) faces of bcc tungsten.
An unusual AFM state was predicted theoretically \cite{Hafner,SB1}
and confirmed experimentally for the bcc-Fe/W(001) system \cite{SB1}
(in fact, the true ground state seems to be  a more complex one
exhibiting a chiral state \cite{chiral}).
There is also evidence  from theory that the ground state
of the Co/W(001) overlayer is AFM, whereas  Mn and
Cr overlayers have a FM ground state \cite{Ferriani}.
This means that a (Fe,Mn)-alloy overlayer on the bcc-W(001) should
exhibit a crossover from an AFM  to a
FM ground state \cite{martin}.
Finally, a similar AFM- to FM-crossover was predicted for
the bcc-Fe/(Ta,W)(001) alloy substrate system  \cite{SB3}, where the
ground state of the bcc-Fe/Ta(001) is  FM. However,
only the leading exchange integrals were estimated in the latter
case \cite{SB3}.

The (001)-faces of the fcc- and bcc-substrates exhibit a simple
rectangular array of lattice sites and differ essentially by the
$d:a_{L}$-ratio, where $a_{L}$ is the lateral lattice constant
 and $d$ is the interlayer distance.
Therefore an investigation of the magnetic properties of
Fe-overlayers  on the (001) faces of fcc  transition metals
poses an interesting problem.
A particularly suitable system is the fcc-Fe/Ir(001) where
very thin overlayers were grown successfully with negligible
Fe-Ir intermixing \cite{halle2}. Furthermore,
there exist reliable LEED measurements \cite{halle2}
elucidating the detailed geometry, while preliminary
MOKE-studies \cite{halle2} indicate no magnetic
signal in the limit of a monolayer coverage;  a situation
similar to the bcc-Fe/W(001).

The clean Ir(001) surface undergoes the $(5\times1)$
quasihexagonal reconstruction.
The finite Fe-coverage (larger than 0.25 monolayer) lifts the
reconstruction \cite{halle2}.
In order to avoid possible corrugation and Fe-Ir intermixing
the metastable unreconstructed $(1\times1)$-Ir(001) surface was
prepared which is, however, stable for room temperatures and
below (for details see Ref.~\onlinecite{halle2}).

The aim of the present study is a comprehensive first-principles
investigation of the properties of an fcc-Fe/Ir(001) monolayer system.
In a first step we perform a structural study of Fe/Ir(001) by using
two highly accurate DFT methods, namely the WIEN2k \cite{wien2k} and VASP
\cite{vasp1,vasp2} codes.
In the next step we investigate the magnetic structure of the geometrically
relaxed system by comparing total energies of the non-magnetic (NM),
FM-, and AFM-configurations ($c$(2$\times $2)-AFM).
Here, the main problem is the large number of possible magnetic configurations
(including the non-collinear ones) to be considered.
Instead of a brute-force search, we evaluate the total energy of the disordered
local moment (DLM) state similarly to Ref.~\onlinecite{SB3}.
The DLM state is a model state with zero total magnetic moment
which results from the disorder of spin orientations of otherwise non-zero
local magnetic moments.\cite{gyorffy}
A strong indicator for a more complex magnetic state of the system
under consideration is a lower total energy of the DLM state
as compared to the NM  and FM states.
One should note, however, that if some specific magnetic state is the ground
state, e.g., the $c$(2$\times $2)-AFM, its total energy is usually lower than
 that of the DLM state.
On the other hand, in random magnetic systems, non-collinear states
can be the ground state as found, e.g. in fcc-NiMn alloys \cite{akai} or in
(Cu,Ni)MnSb Heusler alloys \cite{cunimnsb}.
The usefulness of the DLM concept was demonstrated recently for both
overlayer studies \cite{SB3} and disordered magnetic semiconductors
\cite{akai2}.
The DLM picture may be straightforwardly implemented in the framework
of the coherent-potential approximation (CPA) \cite{akai2}.
Therefore in a third step,
we  perform  studies based on the Green function
implementation of the TB-LMTO-CPA method \cite{book} in the framework of
the surface Green function (SGF) approach, in addition to the above mentioned
collinear WIEN2k and VASP calculations.
The TB-LMTO-SGF approach employs a realistic semiinfinite sample
geometry (no slabs or periodic supercells) and allows to implement the DLM model.
The one-electron potentials are treated within the atomic sphere approximation (ASA);
the dipole barrier due to the sample electrons in the vacuum is
included in the formalism.
TB-LMTO-SGF even allows to include the effect of layer relaxations
\cite{SB3}, provided, they are known either from full-potential
calculations or from experiment.
An important advantage of the TB-LMTO-SGF approach is the possibility to
estimate exchange interactions between magnetic atoms in the overlayer
by a straightforward generalization of the well-approved bulk concept
\cite{Lie,eirev}.
Summarizing, the TB-LMTO-SGF approach is a very useful tool for a qualitative understanding
of the results while the full-potential approaches are superior concerning
the quantitative values and thus provide so to say the corner-stones.
We will demonstrate that at least in the present case the results of
both types of calculations are in a good quantitative agreement
which justifies our assumptions and reinforces our conclusions.

\section{Computational details}
First principle density functional theory calculations were performed
using both the all-electron full potential linearized augmented plane wave (FP~LAPW)
code WIEN2k and the Vienna ab initio simulation package VASP \cite {vasp1}, using the projector
augmented wave scheme \cite {vasp2}. In the FP~LAPW calculations the Fe/Ir(001) systems
were modeled by the eleven-layer repeated slab with  $\approx $ 10~\AA ~vacuum in between.
All slabs were symmetric with respect to the middle layer. We allowed the relaxations for top three layers,
the remaining interlayer distances were fixed to the bulk values of 1.92~\AA.
For VASP  repeated asymmetric slabs with seven layers Ir and a single  Fe mono-layer on
one side and also symmetric slabs with eleven substrate layers and  Fe mono-layers on both sides
were used, which were separated by  at least 19~\AA ~vacuum. All layer distances have been
 relaxed, and turned out to be essentially the same for both setups.
Two DFT potential  approximations have been employed: the local density approximation
\cite{pw92} (LDA) and the generalized gradient approximation\cite{pbe} (GGA) for WIEN2k; the
GGA according to Perdew and Wang (PW91) \cite {pw91} as well as  the
LDA as given by Perdew-Zunger\cite{pzca} (Ceperly-Alder)\cite{cepa} for VASP.
We have tested NM-, FM-, and $c$(2$\times $2)-AFM magnetic arrangements,
all performed in the $c$(2$\times $2)-structure for a reliable comparison of total energies.
Technically, we have used a Brillouin zone  sampling with 21-36 special k- points in the irreducible
 two-dimensional wedge. The difference between input and output charge
density in the final iteration was better than 0.1 me a.u.$^{-3}$.  The total force on single atoms
was in every case smaller than 
1mRy/bohr.

In all TB-LMTO-SGF calculations the LDA approximation and experimental layer relaxations
\cite{halle2} were used. The vacuum above the overlayer was simulated as usual by empty
 spheres (ES).  Electronic relaxations were allowed in
four empty spheres  adjoining the overlayer, the overlayer itself, and
in five adjoining Ir substrate layers.
This finite system was sandwiched selfconsistently between an frozen
semiinfinite fcc-Ir(001) bulk  and the ES vacuum-space including the dipole surface barrier.
Additionally to  NM-, \mbox{FM-} and $c$(2$\times $2)-AFM configurations,
we have studied DLM-arrangements.

In the framework
of the TB-LMTO-SGF method the exchange integrals $J^{\rm Fe,Fe}_{i,j}$
between sites $i,j$ in the magnetic overlayer may be expressed
as follows \cite{eirev,Lie}
\begin{equation}
J^{\rm Fe,Fe}_{i,j}  =
\frac{1}{4 \pi} \, {\rm Im}
\int_{C} \, {\rm tr}_L
\left[ {\Delta^{\rm Fe}_{i}(z)} \,
g^{\uparrow}_{i,j}(z) \,
{\Delta^{\rm Fe}_{j}(z)} \,
g^{\downarrow}_{j,i}(z)
\right] \, {\rm d} z \, .
\label{eqJ}
\end{equation}
Here, the trace extends over $s-,p-,d-,f$-basis set, the quantities
$\Delta^{\rm Fe}_{i}$ are proportional to the calculated exchange
splittings, and the Green function $g^{\sigma}_{i,j}$ describes the
propagation of electrons of a given spin ($\sigma=\uparrow,\downarrow$)
between sites $i,j$.
It should be noted that both the direct propagation of electrons in
the magnetic overlayer and the indirect one via the Ir-substrate are
included in Eq.~(\ref{eqJ}) on an equal footing.
Finally, the energy integration extends over all occupied valence
states up the Fermi energy $E_{F}$ which is technically performed
by integrating over the contour $C$ in the complex energy plane.
For more details see Ref.~\onlinecite{eirev}.
Once the exchange interactions were known, we constructed a two-dimensional (2D)
classical Heisenberg Hamiltonian to describe the magnetic behavior of the
Fe-overlayer on a non-magnetic fcc-Ir(001) substrate
\begin{equation}
H = -\ \sum_{i \neq j} J^{\rm Fe,Fe}_{ij}\ {\bf e}_i
\cdot {\bf e}_j \, .
\label{eqH}
\end{equation}
In Eq.~(\ref{eqH}), ${\bf e}_i$ denotes the orientation of the
Fe-magnetic moment at the site $i$.
By construction, the value of the corresponding magnetic moment is included
in the definition of $J^{\rm Fe,Fe}_{ij}$, and  positive
(negative) values denote  FM (AFM) couplings.
An early study of exchange interactions in fcc-Fe,Co/Cu(001) systems
is found in Ref.~\onlinecite{pajda}, whereas some recent estimates of
exchange integrals for a bcc-Fe/W(001) overlayer were obtained either by
a supercell approach \cite{SB3,leonid} or by an approach closely
related to the present one \cite{szunyogh,shick}.

A Green-function approach like the present one to calculate exchange
interactions  has a particular advantage over a supercell approach \cite{SB3,leonid}:
Exchange interactions can be evaluated easily and reliably
even for disordered overlayers and partial coverages.

\section{Results and discussion}

\subsection{Structure and magnetism}

The results of the total energy calculations are summarized in
Tables I-IV.

\begin{table}[h]
\caption{Calculated ( LDA/GGA,  WIEN and VASP, respectively)
and experimental \cite{halle2} (LEED) interlayer distances $d_{ij}$
between top three sample layers (1-Fe overlayer, 2-top Ir layer,
3-second Ir layer) for fcc-Fe/Ir(001) in the nonmagnetic (NM), ferromagnetic
(FM), and $c$(2$\times $2)-antiferromagnetic (AFM) states. H$_b$-AFM
displays the influence of  0.5 ML hydrogen adsorbed on favourable bridge positions.
The interlayer (001)-distance in the bulk iridium is 1.92 \AA.
}
\label {geo}
\begin{tabular}{c|c|c|c|c|c|c|c}
\hline
\multicolumn{2}{c|}{\raisebox {-1.0ex} [0.0ex] [-1.0ex] {$d_{ij}$}}& \multicolumn{2}{c}{ $d_{12}$ [\AA ]} & \multicolumn{2}{|c}{ $d_{23}$ [\AA ]} & \multicolumn{2}{|c}{ $d_{34}$ [\AA ] }\\
\cline{3-8}
\multicolumn{2}{c|}{}& LDA & GGA & LDA & GGA & LDA & GGA\\
\hline
\raisebox {-1.0ex} [0.0ex] [-1.0ex] {NM}&WIEN&1.52  & 1.61 & 1.95 & 2.00 & 1.86 &1.93\\
&VASP& 1.51 & 1.58 & 1.99 & 2.05 & 1.88 &1.94\\
\hline
\raisebox {-1.0ex} [0.0ex] [-1.0ex] {FM}&WIEN&1.64&1.78&1.91&1.95&1.88&1.94\\
&VASP&1.60&1.76&1.94&1.98&1.89&1.97\\
\hline
\raisebox {-1.0ex} [0.0ex] [-1.0ex] {AFM}&WIEN&1.59&1.69&1.93&1.98&1.88&1.93\\
&VASP&1.55&1.66&1.97&2.02&1.88&1.94\\
\hline
H$_b$-AFM&VASP&1.58&1.67&1.97&2.01&1.88&1.94\\
\hline
LEED&&\multicolumn{2}{c|}{1.69}&\multicolumn{2}{c|}{1.96}&\multicolumn{2}{c}{1.91}\\
\hline
\end{tabular}
\end{table}
In Table I we present results of the structural minimization and
compare our results with experiment.
We found that the ground state is antiferromagnetic and that the
interlayer distances obtained
for this order in GGA approximation agree very well with the results
of LEED structure analysis\cite{halle2}.
For all calculations we used the experimental lattice constant a=3.84~\AA
~in layers, which lies
between the calculated LDA bulk value (3.81~\AA) below
and the GGA (3.87~\AA) above.
For this reason, the calculated substrate interlayer distances are
slightly underestimated in LDA ($\approx$~0.04~\AA) and  overestimated in GGA
 ($\approx$~0.03~\AA),
as the system tries to keep its respective equilibrium volume.  The magnetic
state influences the equilibrium surface geometry considerably at
the Fe/Ir interface, while the changes in the substrate are less pronounced.
A possible contamination with hydrogen, due to the preparation process should
be hard to detect, as the hydrogen induced changes in the spacings are rather
small and around the experimental error limit.
The differences for first interlayer  distance between
the FM and $c$(2$\times $2)-AFM configurations amount to
$\approx $~0.10~\AA, while the NM Fe/Ir spacing is even smaller by 0.08 \AA.
The calculated top layer
relaxations (about 12\%) is smaller than the one obtained for
the similar bcc-Fe/W(001) system (about 14-19~\%) \cite{Ferriani}.

The first Ir-Ir distance is  slightly expanded with respect to its bulk value,
where one has to keep in mind that the bulk spacing is enhanced for GGA
and decreased for LDA. The next spacing is reduced leading to a oscillatory
pattern of interlayer distances found in many metallic systems.

\begin{table}
\caption{Calculated work functions $\Phi $ in eV
for Ir(001) (neglecting a possible lateral reconstruction) and of
Fe/Ir(001) in various magnetic states.
Symbols Ir, NM, FM, AFM,  DLM H$_b$-AFM, and H$_b$-FM
denote respectively, the Ir(001) surface,
and nonmagnetic, ferromagnetic, $c$(2$\times $2)-antiferromagnetic,
and disordered local moment states of the Fe/Ir(001) overlayer, as well as
the FM and AFM states with  0.5 ML hydrogen adsorbed on
bridge positions.
The values correspond to the respective calculated relaxed geometries (see Table  \ref {geo})
for  WIEN/VASP,
and to the experimental one \cite{halle2} for LMTO.
The experimental value for the Ir(001) surface is 
5.67 eV\cite{skriver}.
}
  \begin{tabular}{cccccccc}
    \hline
    $\Phi $[eV] & \rm{Ir} & \rm{NM} & \rm{FM} & \rm{AFM} & \rm{DLM} & \rm{H$_b$-FM} & \rm{H$_b$-AFM} \\
    \hline
   \rm{WIEN-GGA}   & 5.65  & 4.86   &  4.38    & 4.45   & $-$ & $-$ & $-$ \\
   \rm{VASP-GGA}   & 5.62  & 4.82   &  4.29    & 4.37   & $-$  & 4.71 & 4.65 \\
  \rm{WIEN-LDA}   & 5.92  & 5.14  &  4.58    & 4.66   & $-$  & $-$ & $-$\\
   \rm{VASP-LDA}   & 5.89  & 5.13   &  4.59    & 4.66   & $-$   & 5.01 & 5.04 \\
   \rm{LMTO-LDA} & 6.22 & 5.05 & 4.79 & 4.76 & 4.67 & $-$ & $-$ \\
    \hline
  \end{tabular}
\end{table}

The calculated work functions are presented in Table II.
The experimental value of 5.7~eV for the fcc-Ir(001) is reasonably
reproduced by the present calculations (no surface reconstruction).
Our calculations show that the iron overlayer reduces the sample
work function by $\approx $ 1~eV (no experimental data have
been found). Hydrogen adsorbed on the overlayer should increase
the work function by $\approx $ 0.4~eV.
The calculated values for the overlayers are close to the work function of
 bcc-Fe (about 4.5~eV) obtained for a polycrystalline sample \cite{skriver}.
While the full potential VASP and WIEN codes agree very well with each other,
the TB-LMTO-SGF approach slightly overestimates the values.

\begin{table}
\caption{Calculated stabilities (in mRy/Fe atom) of the various magnetic phases of Fe/Ir(001) as
obtained by the WIEN, VASP, and LMTO codes for the unrelaxed geometry
($d_{\rm Fe-Ir}$=1.92~\AA). The ferromagnetic ground
state has the lowest energy and serves as the point of  reference.
}
\label {stab_unrelaxed}
  \begin{tabular}{cccccccc}
    \hline
     $\Delta $[mRy]& \rm{NM$_{\rm LDA}$} &   \rm{AFM$_{\rm LDA}$}
     & \rm{DLM$_{\rm LDA}$} & \rm{NM$_{\rm GGA}$} & \rm{AFM$_{\rm GGA}$}
              \\
    \hline
   \rm{WIEN}        & 49.1 & 2.2 & - & 62.7 & 2.2  \\
   \rm{VASP}        & 44.5 & 3.8 & - & 60.4 & 3.9 \\
   \rm{LMTO} & 42.5&  5.1  & 2.1 & $-$&$-$ \\
    \hline
  \end{tabular}
\end{table}

\begin{table}
\caption{Calculated stabilities (in mRy/Fe atom) of the various magnetic phases of Fe/Ir(001) as
obtained by the WIEN, VASP, and LMTO codes for relaxed geometries.
The stabilities correspond to the respective calculated relaxed geometries (see Table  \ref {geo})
for  WIEN/VASP, and to the experimental one \cite{halle2} for LMTO.
The antiferromagnetic ground
state has the lowest energy and serves as the point of reference.
}
\label {stab_relaxed}
  \begin{tabular}{cccccccc}
    \hline
    $\Delta $[mRy] & \rm{NM$_{\rm LDA}$} &  \rm{FM$_{\rm LDA}$}& \rm{DLM$_{\rm LDA}$}&
              \rm{NM$_{\rm GGA}$} & \rm{FM$_{\rm GGA}$}
               \\
    \hline
   \rm{WIEN}        & 25.8 &  7.8 &  - &  38.4 & 5.3  \\
   \rm{VASP}        & 22.1 & 8.6 & - & 38.7 & 5.5  \\
   \rm{LMTO} & 27.7&  5.0  &   0.8 & $-$ & $-$ \\
    \hline
  \end{tabular}
\end{table}

\begin{table}
\caption{Influence of 0.5 ML adsorbed hydrogen on the calculated stabilities (in mRy) of the
 various magnetic phases of Fe/Ir(001) as obtained by  VASP for relaxed geometries
 (see Table  \ref {geo}).
The antiferromagnetic ground
state has the lowest energy and serves as the point of  reference.
}

\label {stab_hydrogen}
  \begin{tabular}{ccccccc}
    \hline
    $\Delta $[mRy] & \rm{NM$_{\rm LDA}$} &  \rm{FM$_{\rm LDA}$}&
              \rm{NM$_{\rm GGA}$} & \rm{FM$_{\rm GGA}$}
               \\
    \hline
   \rm{clean}        & 22.1 &  8.6 &   38.7 & 5.5  \\
   \rm{bridge-H}        & 14.4 & 3.7  &  29.5 & 2.4  \\
    \hline
  \end{tabular}
\end{table}

The results of magnetic stability calculations are presented in
Tables III - V.
Different theoretical approaches were used and compared in Tables
for the unrelaxed geometry (Table III) as well as for the realistic,
relaxed case (Table IV) including the possibility of residual adsorbed hydrogen
(Table V).
It should be noted that one may only compare different LDA or GGA energies
directly to each to other.
Our calculations clearly show, that the nonmagnetic case can be safely excluded.
All models with local magnetic moments have a substantially lower
total energy.
The most striking result, obtained by all methods, is the fact that
while the FM is the ground state for an unrelaxed geometry, the layer
relaxations stabilize the $c$(2$\times $2)-AFM phase.
This is in a striking contrast to the closely related bcc-Fe/W(001)
case \cite{Hafner,SB1} where the antiferromagnetism of
bcc-Fe/W(001) is robust with respect to the structural relaxations.
Interestingly adsorbed hydrogen  also does not change the picture as
evident from Table V. Differences only get smaller,  but the general trend is preserved.
However, strictly speaking, the $c$(2$\times $2)-AFM may not be the true
ground state of fcc-Fe/Ir(001). Similar to bcc-Fe/W(001) one
should consider other possibilities, e.g., the $p$(2$\times $1)-AFM
or even some non-collinear configurations, such as spin-spirals
or a chiral state \cite{chiral}.
To shed some light on this issue we included in Tables~III and IV the
results of  DLM calculations as performed in the framework of the
TB-LMTO-SGF approach.
The sufficient reliability of the TB-LMTO-SGF approach for the present purpose
is confirmed by a comparison with accurate full
potential calculations for  NM-, FM-, and $c$(2$\times $2)-AFM
configurations in both the ideal and relaxed geometries.
This also justifies the use of the TB-LMTO-SGF approach to
obtain exchange interactions in the Fe-overlayer in the following.
It is quite obvious that, if the DLM state is the ground state compared to the
NM- and FM-states, then a more complex, AFM-like state can exist and
this conclusion can be reached without performing calculations for many
possible candidates.
But clearly, this fact does not render the necessity of searching
for the true ground state of the system obsolete, but rather represents
a reliable qualitative indicator for a more complex magnetic state
of the system.

One can speculate about the structural origin of such an AFM order.
As already mentioned, both bcc(001) and fcc(001) have a common
square-lattice structure of magnetic atoms which differ
 by the ratio $d$~:~$a_{L}$, where $a_{L}$ is the lateral
lattice constant and $d$ is the layer spacing.
 For bcc(001) we have $d=a_{L}/2=a/2$ (where $a$ is the bulk lattice
constant) while for fcc(001) $d=a/2$ and $a_{L}=a/\sqrt2$ leading
to a larger $d$~:~$a_{L}$.
The reduction of the $d$~:~$a_{L}$ ratio stabilizes the
AFM-/DLM-state for this fcc(001) surface. This ratio is sufficiently small for comparable bcc(001) surfaces even in the unrelaxed geometry (see e.g. Fe/W(001) \cite{SB1}).
It is well-known that the exchange interaction between Mn spins becomes
antiferromagnetic for smaller distances, a trend we observe here
for exchange interactions in the Fe-overlayer as detailed below.
This is a first strong indication that  indirect interactions of
Fe-spins via the Ir-substrate play an essential role for the
fcc-Fe/Ir(001) magnetism.
A dominant character of indirect interactions as compared to direct
ones between Fe-spins in the overlayer will result in a strong
dependence of exchange interactions on the Fe-Ir interlayer distance
as we shall see below.

\subsection{Densities of states and exchange interactions}

\begin{figure}[h]
\center \includegraphics[width=8.5cm]{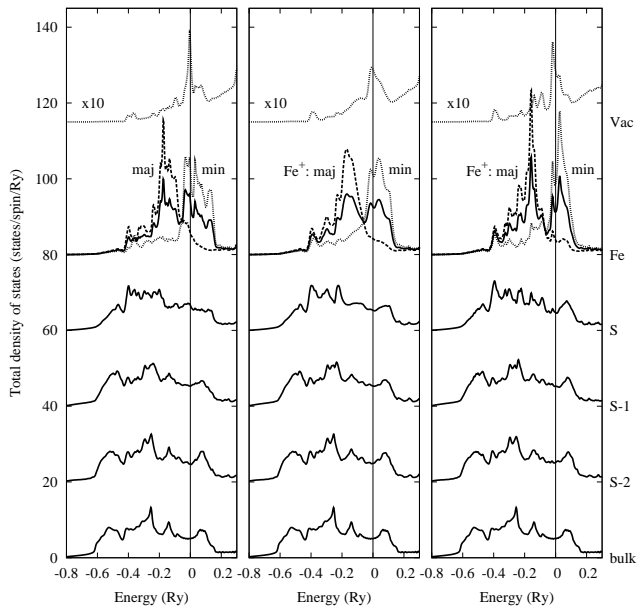}
\caption {Total layer-resolved densities of states (DOS) for
the fcc-Fe/Ir(001) overlayer and experimental layer relaxations
\cite{halle2} based on the LMTO approach. In the case of
the Fe-overlayer the total DOSs are additionally split into majority
(dashed lines) and minority (dotted lines) contributions:
(a) the FM case, (b) the DLM case, and (c) the $c$(2$\times $2)-AFM
case. Only one spin-orientation (denoted as Fe$^{+}$) is plotted
for the DLM and AFM cases. Symbols Vac, Fe, S-1, S-2, and bulk
denote the first vacuum layer, Fe-overlayer, first  and second substrate
Ir-layers, and the fcc bulk Ir host DOSs, respectively. The Fermi
energy is shifted to the energy zero.
}
\label{fig1}
\end{figure}

In Fig~\ref{fig1} we present the  layer-resolved densities of states
(DOS) for the various magnetic states.
Compared to the Ir bulk, the most important feature observed
for all cases is an extra
contribution to the overlayer DOS around the Fermi energy
due to the minority Fe-states.
The large exchange splitting of majority and minority Fe-states
is due to the enhanced overlayer magnetic moment (about 2.65~$\mu_{B}$
in all cases) which illustrates the rigidity of the Fe-moment with
respect to changing spin-orientations.
The large Fe moment is due to the large lateral overlayer
lattice constant as given by the Ir-substrate as well as to the reduction
of coordination number at the surface typical for overlayer systems.
There is also a relevant extra peak in the DOS in the vacuum close to the
sample surface which gives a possibility to detect it in  STM
measurements.
We also observe a strong reduction of the Fe-overlayer imprint
on the deeper Ir-substrate layers: already the second
Ir-substrate layer is almost bulk like.
There is a small induced moment on the first-substrate Ir-layer
of the order of 0.1~$\mu_{B}$ while other induced moments are
strongly damped in an oscillatory manner (the Friedel-like
oscillations) into the Ir-substrates and their values are
of the order of 0.01~$\mu_{B}$ and smaller.
It should be noted that induced moments in the Ir-substrate and
in the vacuum are much smaller and more strongly damped in both
substrate and vacuum for the $c$(2$\times $2)-AFM case as compared to
the FM case, while they even collapse to zero for the DLM state.
The induced moment in the substrate/vacuum are more than two-times
smaller as those in the bcc-Fe/W(001) \cite{SB3}.
The above results are confirmed by full-potential slab-model calculations using both
WIEN and VASP codes.

\begin{figure}
\center \includegraphics[width=8.8cm]{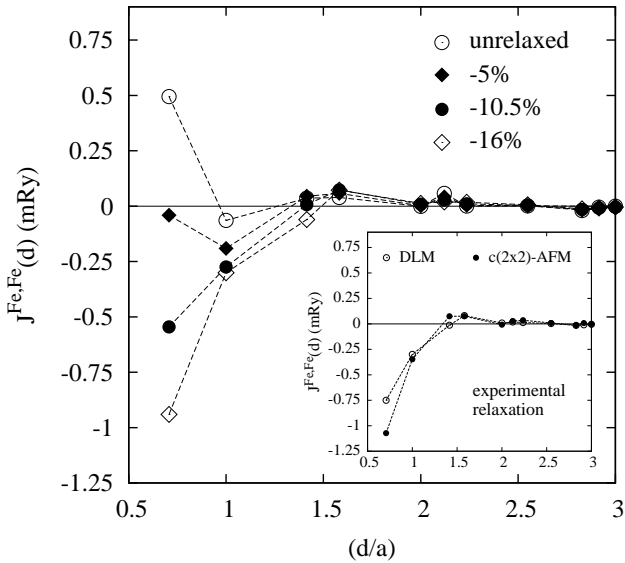}
\caption {Exchange interactions among Fe-atoms in the Fe/Ir(001)
overlayer for the unrelaxed case and model layer relaxations
evaluated as a function of the reduced interatomic distance ($d/a$),
where $a$ denotes the lattice constant.
Numbers attached to symbols indicate the reduction of the Fe-Ir
interlayer distances in \% as compared to the bulk value of 1.92~\AA.
The inset shows exchange interactions for AFM and DLM state
for the experimental layer relaxations\cite{halle2}.
 All results were obtained assuming the DLM-reference state.
}
\label{fig2}
\end{figure}

Exchange interactions $J^{\rm Fe,Fe}(d)$ have been determined
by the TB-LMTO-SGF method for Fe/Ir(001) as a function of the
interatomic Fe-Fe distance $d$ for both unrelaxed and relaxed
geometries, and FM-, DLM-, and AFM-reference states.
Additionally we present results in the DLM state for a simple
layer-relaxation model where only the Fe-Ir interlayer distance
is reduced from the unrelaxed value (1.92~\AA) to values of
1.82~\AA, 1.72~\AA, and 1.62~\AA ~ corresponding to a reduction
of 5~\%, 10.5~\%, and 16~\%.
These results are shown in Fig.~\ref{fig2}.

Calculated exchange integrals are rather similar, e.g., all
leading to the AFM interactions for the relaxed case irrespective of the reference state.
The problem of the choice of reference state for estimate of
exchange integrals in the present context was also addressed
in Ref.~\onlinecite{szuny}.
We have chosen the DLM reference state (with self-consistently
calculated spin-polarized potentials and with corresponding
modification of Eq.~(\ref{eqJ})) for the magnetic stability study
below as it assumes no magnetic ordering and for our purposes
is thus most suitable. We admit that to estimate e.g. the critical
temperatures other choices may be more suitable.

With increasing layer relaxations we observe a clear tendency
towards dominating AFM interactions which stabilize the AFM-like
state in the overlayer.
The dominating role of indirect interactions between Fe-atoms
via the Ir-substrate is obvious: the only varying quantity is
the Fe-Ir distance and thus the Fe-Ir hybridization.
Results for experimental layer relaxations (inset in Fig.~\ref{fig2})
are between model cases of 1.72~\AA, and 1.62~\AA.
The fact that strong AFM coupling in the layer-relaxed case were
obtained from a reference FM state indicates the robustness of
the AFM order (more precisely, of a more complex magnetic state)
for the relaxed Fe/Ir(001) overlayer.

\begin{figure}
\center \includegraphics[width=7.8cm]{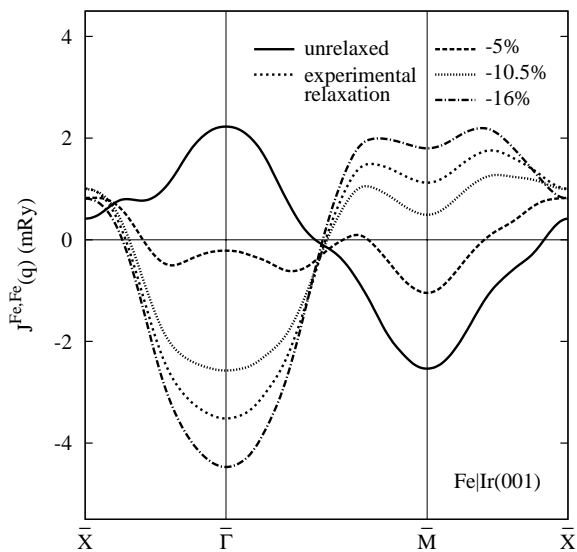}
\caption {Lattice Fourier transformation of the real-space exchange
interactions J$^{\rm Fe,Fe}_{ij}$, J({\bf q$_{\|}$}), for the ideal
unrelaxed geometry, the experimental layer relaxations \cite{halle2}
as well as for three model Fe-Ir layer relaxations (same as in
Fig.~\ref{fig2}) which were obtained for the reference DLM state of
the Fe/Ir(001) overlayer.
Here, {\bf q$_{\|}$}=$\bar{\rm X}$=$2\pi /a_L(1/2,1/2)$,
{\bf q$_{\|}$}=$\bar{\Gamma }$=$2\pi /a_L(0,0)$, and
{\bf q$_{\|}$}=$\bar{\rm M}$=$2\pi /a_L(1,0)$.
}
\label{fig3}
\end{figure}

To further investigate this point we present in Fig.~\ref{fig3} the
result of the lattice Fourier transformation of exchange integrals
(\ref{eqJ}) as obtained for the DLM reference state.
It should be noted that the DLM state is neither the ground state
for the ideal geometry nor for the experimental layer-relaxed model.
On the other hand, possible magnetic phases of the 2D-Heisenberg
model (\ref{eqH}) can be obtained by studying its stability with
respect to the periodic excitations.
A similar approach was successfully used in the study of the complex
magnetic stability of bcc-Eu: starting from the FM reference
state, a proper spin-spiral ground state was obtained in good
agreement with the experiment \cite{bcceu}.
Due to the sign convention in eqaution~(\ref{eqH}) the maximum of
J({\bf q$_{\|}$}) corresponds to the ground state (the energy
minimum).
It is obvious that the ground state for the unrelaxed model is the
ferromagnetic state (the maximum of J({\bf q$_{\|}$}) is obtained
for {\bf q$_{\|}$} = $2\pi /a_L(0,0)$.
With reduced Fe-Ir interlayer distance we observe a quick decrease
of the stability of the FM state and the ground state is found
for the ordering vector {\bf q$_{\|}$}=$2\pi /a_L(1/2,1/2)$ which
corresponds to the $c$(2$\times $2)-AFM state (for the Fe-Ir interlayer
distance $d$=1.82~\AA ~or reduced by 5\%).
If the Fe-Ir interlayer distance further decreases (by 10.5\% and
16\%) the stability of the FM state further decreases and a new,
complex spin-spiral like magnetic state becomes more stable
as compared to the $c$(2$\times $2)-AFM state (ordering vector on the
line $\bar{\rm X}$-$\bar{\rm M}$ in the irreducible surface Brillouin zone).
On the other hand, the $c$(2$\times $2)-AFM state becomes more and
more stable as compared to the FM state in accordance with
the total energy calculations.
Also, both for the experimental layer-relaxation model \cite{halle2}
and for the model with the largest reduction of the Fe-Ir interlayer
distance (by 16\%) the $p$(2$\times $1)-AFM state (the ordering vector
{\bf q$_{\|}$}=$2\pi /a_L(1,0)$) has lower energy as compared to the
FM- and $c$(2$\times $2)-AFM states.

In order to advance the understanding of the present substrate induced coupling
a comparison with an otherwise similar 4$d$ substrate would be beneficial.
Rhodium crystallizes also in the fcc structure, has the
same number of valence
electrons (9) as iridium and a similar lattice constant (3.80~\AA),
while the value of the work function, 5.11~eV \cite{rh}, is smaller as well as
the spatial extent of the 4$d$-wave functions of Rh as
compared to 5$d$-wave functions of Ir.
The results of a similar study of  the magnetic stability for fcc-Fe/Rh(001)
overlayer as a function of the Fe-Rh interlayer distance will be presented in the following.
The same relative interlayer reductions as for fcc-Fe/Ir(001),
namely by 5~\%, 10.5~\%, and 16~\% will be used.

\begin{figure}
\center \includegraphics[width=7.8cm]{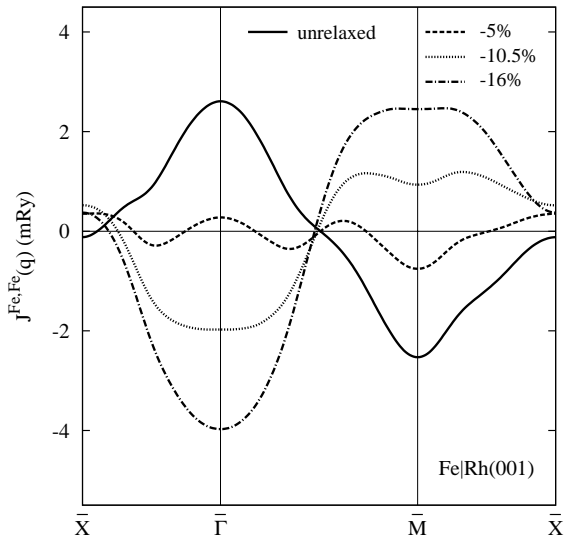}
\caption {Lattice Fourier transformation of real-space exchange
interactions $J^{\rm Fe,Fe}_{ij}$, $J$({\bf q$_{\|}$}), for the
unrelaxed geometry, and three model Fe-Rh interlayer relaxations for the
fcc-Fe/Rh(001) system. The identical relative interlayer distances were used
for both the fcc-Fe/Ir(001) system and for the DLM reference state.
}
\label{fig4}
\end{figure}

The results are summarized in Fig.~\ref{fig4}.
The FM state is again the ground state for the unrelaxed case, and
we observe a similar Fe-Rh interlayer distance reduction effect
on the magnetic stability as for fcc-Fe/Ir(001).
The reduction of the Fe-Rh interlayer distance by about 5~\%
seems to be the point where the FM state is no longer the
ground state and the $c$(2$\times $2)-AFM state is stabilized (the ordering
vector {\bf q$_{\|}$}=$2\pi /a_L(1/2,1/2))$.
However, with increasing reduction of the Fe-Rh  distance the more complex
AFM state is stabilized until for an interlayer reduction of 16~\%  this state become the ground state
(see an indication of this case in the Fig. 4: maximum in the
neighborhood of the {\bf q$_{\|}$}=$2\pi /a_L(1,0)$ ordering vector representing the
$p$(2$\times $1)-AFM state).
Again the AFM state is stabilized but for a relatively larger reduction
of the Fe-Rh interlayer distance as compared to fcc-Fe/Ir(001), but
otherwise there is a large similarity between the two overlayer systems.
The larger reduction is in agreement with previous total energy study
\cite{Hwang}.

\section{Conclusions}

We have investigated the experimentally prepared Fe/Ir(001) system by a
combination of different  first-principles methods for both supercell slab geometries
(WIEN and VASP codes) and semi-infinite boundary conditions (TB-LMTO-SGF codes).
Using the latter approach we  investigated the magnetic
phase stability of the system by calculating the exchange
interactions between Fe-atoms in the overlayer.
The following conclusions can be drawn:
(i) Calculated relaxed geometries agree well with available experimental
data. A~better agreement is obtained for the Fe-Ir distance
using  GGA rather than LDA potentials. The calculated  geometries 
depend on the
magnetic state (non-magnetic, FM, and AFM states) but the most important
result is the reduction of the Fe-Ir interlayer distance as compared
to the bulk value by about 12~\% for the AFM order.
A possible residual H contamination on the overlayer has practically
no effect on the distances;
(ii) The local Fe-moment is enhanced to about 2.65~$\mu_{B}$
 as compared to its canonical value of 2.15~$\mu_{B}$ in the bcc Fe-metal.
The enhancement is due to both the enlarged lateral lattice constant of
the Fe-overlayer on fcc-Ir(001) and the reduction of nearest-neighbors
there.
The work function of the system with Fe-overlayer is reduced more than 1~eV
as compared to the value found for pure Ir-surface. Hydrogen on the overlayer
increases the work function by $\approx$~0.4~eV;
(iii) Both full potential supercell methods found
 the $c$(2$\times $2)-AFM state to be stable as compared to
the non-magnetic and ferromagnetic states for the correctly relaxed structure.
On the other hand, they find an FM  ground state for a unrelaxed geometry.
Hence, the stability of the AFM state is induced by the substrate
via  layer-relaxations.
The above results were confirmed both qualitatively and quantitatively
by the TB-LMTO-SGF approach assuming a semiinfinite sample geometry
and experimental layer relaxations as obtained in LEED;
(iv) The related fcc-Fe/Rh(001) system behaves similarly to
fcc-Fe/Ir(001), although the tendency towards the formation of the AFM
state upon reduction of the Fe-Rh interlayer distance
is weaker. We found an indication that a $p$(2$\times $1)-AFM state
is stabilized for larger reductions of Fe-Rh interlayer distances;
(v) A detailed study of the magnetic stability based on the
2D-Heisenberg Hamiltonian derived from the first-principles total
energies confirms the stability of the $c$(2$\times $2)-AFM state as compared
to the FM for relaxed model but indicates that more complex,
spin-spiral like state can be stabilized by reducing the Fe-Ir
interlayer distance. Since Ir is a heavy element, the spin-orbit
interaction can have non-negligible effect on the exchange interactions
(in analogy to the Mn/W case\cite{chiral}), and can lead to chiral
magnetic order induced by Dzyaloshinskii-Moriya interaction; and
(vi) We have shown that increasing Fe-Ir hybridization stabilizes
the AFM state. This seems to be a robust effect to warrant similar
stabilization on a much more complex reconstructed Ir(001) surface.
However, such a study goes beyond the subject of the present paper.

Experimentally no-magnetization was found for iron overlayers
thinner than $\approx $ 4 monolayers on fcc-Ir(001) \cite{halle2}
and $\approx $ 6 monolayers on fcc-Rh(001)\cite{Hwang}.
Our study indicates that this is not the result due to disappearing of
the local iron magnetic moments in the top surface layers but rather
due to a more complex "antiferromagnetic-like" order in the
fcc-iron overlayer.

\section*{Acknowledgements}
This work has been done within the project AV0Z1-010-0520 and AV0Z2-041-0507 of the
ASCR. The authors acknowledge fruitful discussions with J. Kirschner,
D. Sander and Z. Tian and the support from the Grant Agency of the
Czech Republic, Contract No. 202/07/0456, 202/09/0775 and COST P19-OC-09028 project.

\end{document}